\documentclass[10pt, twocolumn, pre, aps, superscriptaddress, showpacs]{revtex4-1}
\usepackage{amsmath, graphicx, subfigure, tikz}
\begin{document}
    \title{Across Dimensions: Two- and Three-Dimensional Phase Transitions\\
    from the Iterative Renormalization-Group Theory of Chains}
    \author{Ibrahim Ke\c{c}o\u{g}lu}
    \affiliation{Department of Physics, Bo\u{g}azi\c{c}i University, Bebek, Istanbul 34342, Turkey}
    \author{A. Nihat Berker}
    \affiliation{Faculty of Engineering and Natural Sciences, Kadir Has University, Cibali, Istanbul 34083, Turkey}
    \affiliation{Department of Physics, Massachusetts Institute of Technology, Cambridge, Massachusetts 02139, USA}

    \begin{abstract}
Sharp two- and three-dimensional phase transitional magnetization
curves are obtained by an iterative renormalization-group coupling
of Ising chains, which are solved exactly. The chains by themselves
do not have a phase transition or non-zero magnetization, but the
method reflects crossover from temperature-like to field-like
renormalization-group flows as the mechanism for the
higher-dimensional phase transitions. The magnetization of each
chain acts, via the interaction constant, as a magnetic field on its
neighboring chains, thus entering its renormalization-group
calculation. The method is highly flexible for wide application.
    \end{abstract}
    \maketitle

\section{Introduction: Connections Across Spatial Dimensions}
It is well-known and quickly shown that one-dimensional models
$(d=1)$ with finite-range interactions are exactly solvable and do
not have a phase transition at non-zero temperature
\cite{Frobenius}. Nevertheless, the phase transitions of the $d>1$
models can be distinctively recovered from the correlations in the
exactly solved $d=1$ chains, as we show in the present study.
Specifically, using the exact renormalization-group solution of the
$d=1$ Ising chain (which at non-zero temperatures has no phase
transition and zero magnetization), the finite-temperature phase
transitions and entire magnetization curves of the $d=2$ and $d=3$
Ising models are recovered distinctively
(Fig. 1).  Our method is an approximation and is not obviously systematically improvable towards the exact results.  The method is general and flexible, and thus can be applied to a wide range of systems, such as with random fields and/or random bonds, more complicated local degrees of freedom such as spin-s Ising or q-state Potts.
\begin{figure}[ht!]
\centering
\includegraphics[scale=0.22]{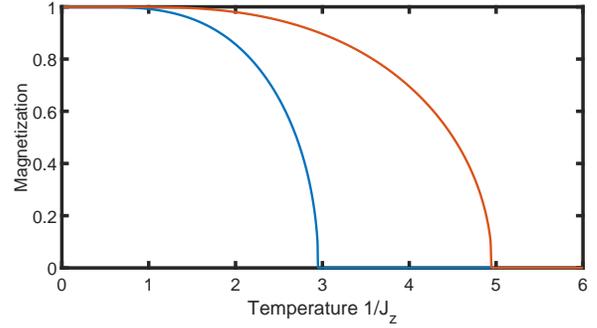}
\caption{From left  to right, magnetizations for $d=2$ and $d=3$
Ising models obtained by coupling exact solutions of Ising chains.
The magnetization of each chain acts, via the coupling constant $J$,
as a magnetic field entering the renormalization-group calculation
of its neighboring chains. This procedure is iterated until the
magnetization curve converges, as seen in Fig. 2, in fact creating
order out of the renormalization-group flows within a disordered phase,
reflecting a crossover from field-like (at lower temperatures) to
temperature-like flows. The magnetization curves on this figure are
actually smooth on reaching zero, but this cannot be seen on the
scale of figure.}
\end{figure}

\begin{figure}[ht!]
\centering
\includegraphics[scale=0.22]{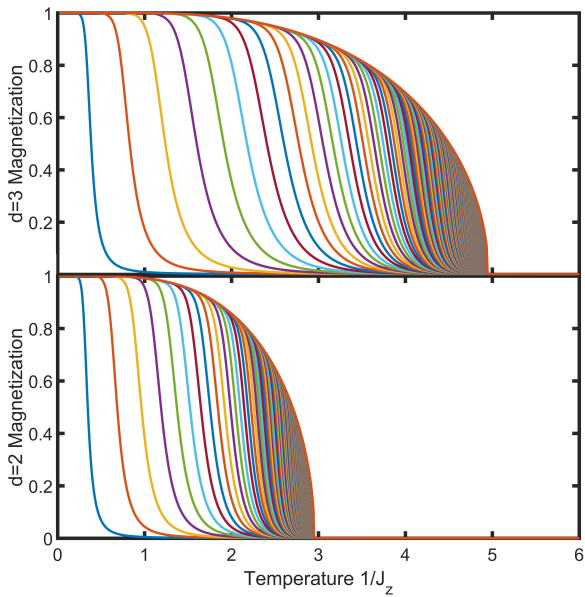}
\caption{Iterations in the magnetization calculations. Starting with
a small seed of M=0.00001 and proceeding with the
renormalization-group calculated magnetizations, after each
iteration, each site applies a magnetic field of $JM$ to its
neighbors on the neighboring chains. The chains are arrayed to give
a $d>1$ dimensional system. The leftmost curve is the result of the
first iteration.  Distinctly for both $d=2$ and 3, the
magnetizations quickly converge (to the rightmost curve) and give
the higher-dimensional phase transitions.}
\end{figure}

The systems that we study are defined by the Hamiltonian
\begin{equation}
-\beta \mathcal{H}=J\sum_{\langle ij \rangle} s_is_j + h\sum_i s_i,
\end{equation}
where at each site $i$, the spin is $s_i=\pm1$ and the first sum is
over all pairs of nearest-neighbor sites $<ij>$.  We obtain the
phase transitions and magnetizations of these Ising systems in
spatial dimensions $d=2,3$ at magnetic field $H=0$, based on the
renormalization-group solution of the $d=1$ system with $H\neq 0$,
as given in Eq.(3).

\begin{figure}[ht!]
\centering
\includegraphics[scale=0.27]{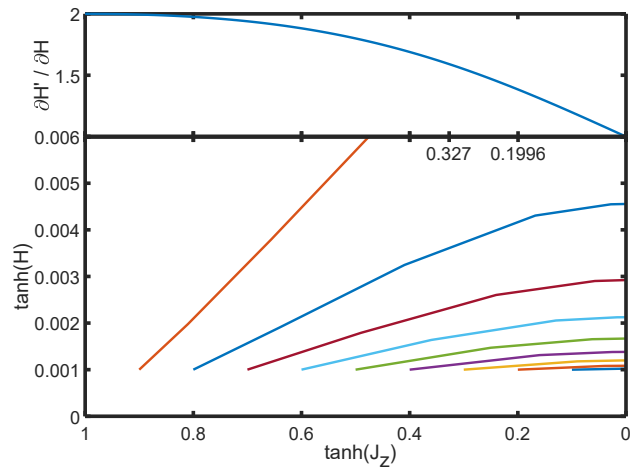}
\caption{Lower panel: Renormalization-group flows (Eq.(2)) of the
$d=1$ Ising model (Eq.(1)).  Trajectories originating at the small
$H=0.001$ and the entire breadth of temperature are given, all
terminating at different locations on the fixed line of the renormalization-group flows at $J_z=0$. Upper
panel: The derivative of the renormalized magnetic field $H'$ with
respect to the unrenormalized $H$, at $H=0$. Its values stay around its maximal value of 2 (at strong coupling the magnetic field of a decimated spin adds to the magnetic field of a remaining spin) until temperatures around
$1/J_z=3$ and crosses to its minimal value of 1 (at weak coupling the magnetic field of a decimated spin does not affect the magnetic field of a remaining spin).  In the lower panel, we compare the magnetic fields acquired by the renormalization-group trajectories originating at $\tanh(J_z)=0.9$ and $\tanh(J_z)=0.1$. The renormalization-group trajectories originating at higher temperatures end on the fixed line at a small value of $H$. This mechanism thwarts the lateral
couplings of the chains and ushers the high-temperature disordered
phase. The calculated transition temperatures for $d=2$ (on left)
and $d=3$ are consistently shown on the middle axis.}
\end{figure}

\section{Renormalization-Group Flows of the $d=1$ Ising Model with Magnetic Field}

The Ising model of Eq. (1) with non-zero magnetic field can be
subjected, in a $d=1$ chain, to exact renormalization-group transformation
\cite{NelsonFisher,Grinstein} by performing the sum over every other
spin (\textit{aka}, decimating, actually a misnomer).

The couplings of the remaining spins (of the thus renormalized
system) are given by the recursion relations:
\begin{multline}
\begin{split}
J'_z =& \frac {1}{4} \ln [R(++)R(--)/R(+-)R(-+)] \,, \\
H' =& \frac {1}{4} \ln [R(++)/R(--)] \,, \\
G' =& b^dG + \frac {1}{4} \ln [R(++)R(+-)R(--)R(-+)] \,, \\
R(\sigma_1&\sigma_3) = \sum_{s_2=\pm1} \exp[-\beta
\mathcal{H}(s_1,s_2)-\beta \mathcal{H}(s_2,s_3)],
\end{split}
\end{multline}
where the primes refer to the quantities of the renormalized system,
$b=2$ is the length rescaling factor, $d=1$ is the dimensionality,
$\sigma_i$ is the sign of $s_i$ and, for calculational convenience,
the Hamiltonian of Eq.(1) has been rewritten in the equivalent form
of
\begin{equation}
-\beta \mathcal{H}=\sum_{\langle ij \rangle}-\beta
\mathcal{H}(s_i,s_j)=\sum_{\langle ij \rangle} [J_z s_is_j + H(s_i+s_j)
+G],
\end{equation}
where $J_z$ is the interaction strength between neighboring spins within the $d=1$ chain.

In the present calculation, the magnetic field $H$ in Eqs.(2,3) represents the lateral interactions to the $d=1$ chain:  Its starting value, before renormalization-group is applied, is
\begin{equation}
H = q J M /2,
\end{equation}
where $q=2,4$ respectively for $d=2,3$ is the number of lateral (off-chain) neighbors of each spin in the chain, $J$ is the lateral interaction, $M$ is the magnetization, and we divide by 2 because each spin gets counted twice in the sum in Eq.(3).  The lateral mean-field approximation of Eq.(4) is applied only at the initial point of the renormalization-group trajectory. This magnetic field $H$ representing the lateral interactions gets renormalized under the renormalization-group transformations, together with the in-chain interaction $J_z$, as given in the recursion relations in Eq.(2).

In Eq.(3), $G$ is the additive constant per bond, unavoidably generated
by the renormalization-group transformation, not entering the
recursion relations as an argument (therefore a captive variable),
but crucial to the calculation of all the thermodynamic densities,
as seen in Sec. III below.\cite{BerkerOstlund,Ilker2,Atalay,Artun}

Typical calculated renormalization-group flows of $(J_z,H)$ are given
in the lower panel of Fig. 3.  All flows are to infinite temperature
$1/J_z = \infty$ (with the exception of the unstable critical fixed
point at zero temperature, zero field $(1/J_z=0, H=0))$.  At infinite
temperature (zero coupling, $J_z=0$) a fixed line occurs in the $H$
direction and is the sink of the disordered phase, which attracts
everything in $(J_z,H)$ except for the single critical point.  However,
we shall see in Sec. IV below that this disordered phase engenders
the ordered phases of $d=2$ and 3.

\begin{figure}[ht!]
\centering
\includegraphics[scale=0.22]{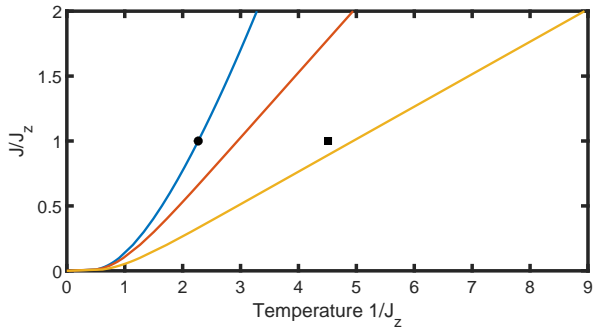}
\caption{Phase boundaries for the isotropic/anisotropic Ising
models: From left to right, the $d=2$ exact boundary
$\exp(-2J)=\tanh(J_z)$ (Ref. \cite{McCoyWu}), the $d=2$ and $d=3$
boundaries calculated by our method.  $J_z$ is the nearest-neighbor
interaction along the chains and $J$ is the nearest-neighbor
interaction lateral to the chains. The exact phase transition points
for the isotropic systems, $J_z=J$, are given by the circle $(d=2)$
and square $(d=3)$ (Ref. \cite{3dIsing, Pelissetto}) data points.}
\end{figure}

The derivatives of the renormalized magnetic field $H'$ with respect
to the unrenormalized $H$, at $H=0$, are shown in the upper panel of
Fig. 3.  Its values stay around its maximal value of 2 (at strong coupling the magnetic field of a decimated spin adds to the magnetic field of a remaining spin) until temperatures around $1/J_z=3$ and crosses to its minimal value of 1 (at weak coupling the magnetic field of a decimated spin does not affect the magnetic field of a remaining spin).  In the lower panel of Fig. 3, we compare the magnetic fields acquired by the renormalization-group trajectories originating at $\tanh(J_z)=0.9$ and $\tanh(J_z)=0.1$. The renormalization-group trajectories originating at higher temperatures end on the fixed line at a small value of $H$.  This mechanism thwarts the lateral couplings of the
chains and ushers the high-temperature disordered phase.

\section{Renormalization-Group Calculation of Thermodynamic Densities}

The thermodynamic densities $\textbf{M} \equiv [1, <s_is_j>,
<(s_i+s_j)>]$, which are the densities conjugate to the interactions
$\mathbf{J_z} \equiv [G, J_z, H]$ of Eq.(3), obey the density recursion
relation
\begin{equation}
\textbf{M} = b^{-d} \textbf{M'} \cdot \textbf{T},
\end{equation}
where the recursion matrix is $\textbf{T} = \partial \mathbf{J_z'} /
\partial\mathbf{J_z}$. Although the renormalization-group recursion relations [Eq.(2)] are nonlinear, this linear equation relating the renormalized and unrenormalized densities is exact.  It is obtained by using the derivative chain rule on $\textbf{M} = (1/N) \partial \ln Z /
\partial\mathbf{J_z}$, where $Z$ is the partition function and $N$ is the number of nearest-neighbor pairs of spins, and is used to calculate densities from renormalization-group theory \cite{BerkerOstlund,Ilker2,Atalay,Artun}, as explained below.

The densities at the starting interactions of the renormalization-group trajectory are calculated by
repeating Eq.(5) until the fixed-line is quasi-reached and applying
the fixed-line densities, variable with respect to the terminus $H$,
on the right side of the repeated Eq.(5):
\begin{equation}
\textbf{M(0)} = b^{-nd} \textbf{M(n)} \cdot \textbf{T(n)}\cdot
\textbf{T(n-1)}...\cdot \textbf{T(1)}\,,
\end{equation}
where \textbf{M(n)} are the densities at the $(J_z,H)$ location of the
trajectory after the $(n)$th renormalization-group transformation
and \textbf{T(n)} is the recursion matrix of the $(n)$th
renormalization-group transformation \cite{BerkerOstlund,Artun}.
Thus, \textbf{M(0)} are the densities of the $(J_z,H)$ location where
the renormalization-group trajectory originates and the aim of the
renormalization-group calculation.  Note that \textbf{M(0)} is
obtained by doing a calculation along the entire length of the
trajectory.  As seen in Fig. 3, the trajectory closely approaches,
after a few renormalization-group transformations, a point $(J_z=0,H)$
on the fixed line and $\textbf{M(n)}\simeq \textbf{M*}(H)$, where
the latter density is calculated on the fixed line.

The densities $\textbf{M*}(H)$ on the fixed line are, by
Eq.(5), the left eigenvector of the recursion matrix
$\textbf{T*}(H)$ at the fixed line with eigenvalue $b^d$.  (Since
the recursion matrix is always non-symmetric, the left and right
eigenvectors are different with the same eigenvalue.)  In the
present case, on the fixed line,
\begin{equation*}
\textbf{T*}(H) =
\begin{bmatrix}
2 & 0 & 2\,\tanh(2H) \\
0 & 0 & 0 \\
0 & \tanh(2H) & 1
\end{bmatrix}\,
\end{equation*}
and the left eigenvector with eigenvalue $b^d=2$ is $\textbf{M*}(H)
= [1, <s_is_j>=(\tanh(2H))^2, <(s_i+s_j)>=2\,\tanh(2H)].$
\begin{figure}[ht!]
\centering
\includegraphics[scale=0.22]{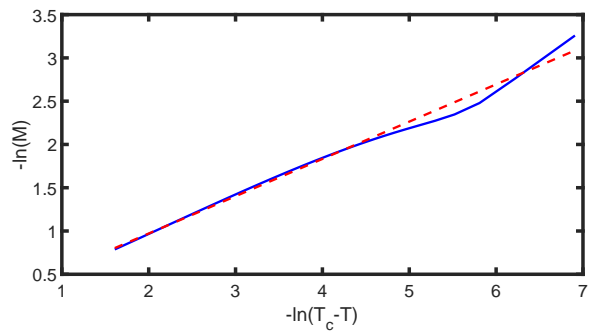}
\caption{Power-law $M \sim (T_C-T)^{\beta}$ fit to our $d=2$ result.
A fit over 6 decades, with a quality of fit $R=99.6$, gives the
critical exponent $\beta = 0.43$, lower than the mean-field value of
$1/2$. However, this value is quite far from the exact value of $\beta= 0.125$.}
\end{figure}
\begin{figure}[ht!]
\centering
\includegraphics[scale=0.22]{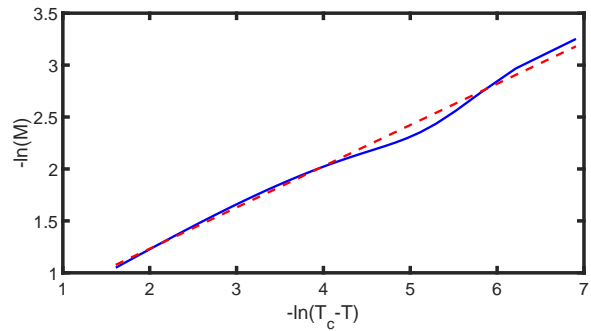}
\caption{Power-law $M \sim (T_C-T)^{\beta}$ fit to our $d=3$ result.
A fit over 6 decades, with a quality of fit $R=99.5$, gives the
critical exponent $\beta = 0.40$, lower than the mean-field value of
$1/2$. However, this value is quite far from the exact value of $\beta= 0.326$.}
\end{figure}

\section{Sharp Magnetization Curves and Phase Diagrams}
The phase diagrams (Fig. 4) for the anisotropic and isotropic Ising
models in $d=2$ and 3 are obtained by repeating our calculation for
different values of the interactions $J_z$ along the chains and $J$
lateral to the chains, and compare well with the exact results also
given in the figure.  Critical exponents are obtained by power-law
$M \sim (T_C-T)^{\beta}$ fitting simultaneously the exponent and the
critical temperature to the curves in Fig. 1. As seen in Figs. 5 and
6, in both cases fitting over 6 decades with a quality of fit of
$R=99.6$ and 99.5, the critical exponents $\beta = 0.43$ and 0.40
are obtained, perhaps meaningfully lower than the mean-field value
of $1/2$.  However, these values are quite far from the exact values of $\beta= 0.125$ and 0.326 in $d=2$ and 3 respectively \cite{McCoyWu,3dIsing,Pelissetto}, which indicates that our approach, although qualitatively not unreasonable and widely applicable, is not that accurate with respect to critical exponents.

\section{Conclusion}

We believe that our method could be easily and widely implemented,
since complex systems (as long as the interactions are non-infinite
ranged) can be solved in $d=1$ \cite{NelsonFisher,Grinstein} and
applied to the higher dimensions as demonstrated here.  Furthermore,
random local densities can be obtained for quenched random systems
\cite{Yesilleten}, in $d=1$ using renormalization-group theory, and
applied with our method to a variety of quenched random systems in
$d>1$. It would also be interesting to apply to systems which show
chaos under direct renormalization-group theory, obtaining an
alternate path to study such chaos \cite{Atalay,Fernandez2,Eldan}.

\begin{acknowledgments}
Support by the Academy of Sciences of Turkey (T\"UBA) is gratefully
acknowledged.
\end{acknowledgments}


\begin{references}

\bibitem{Frobenius} S. B. Frobenius, Preuss. Akad. Wiss. 471 (1908); 514 (1909).

\bibitem{NelsonFisher} D. R. Nelson and M. E. Fisher, Soluble Renormalization Groups and Scaling Fields for Low-Dimensional Ising Systems, Ann. Phys. (N.Y.) {\bf 91}, 226 (1975).

\bibitem{Grinstein} G. Grinstein, A.N. Berker, J. Chalupa, and M. Wortis, Exact renormalization group with Griffiths singularities and spin-glass behavior: The random Ising chain, Phys. Rev. Lett. {\bf
36}, 1508 (1976).

\bibitem{BerkerOstlund} A. N. Berker and S. Ostlund, Renormalisation-group calculations of finite systems: Order parameter and specific heat for epitaxial
ordering, J. Phys. C {\bf 12}, 4961 (1979).

\bibitem{Atalay} B. Atalay and A.N. Berker, A lower lower-critical spin-glass dimension from quenched mixed-spatial-dimensional spin glasses, Phys. Rev. E {\bf 98}, 042125 (2018).

\bibitem{Ilker2} E. Ilker and A. N. Berker, Overfrustrated and underfrustrated spin glasses in d=3 and 2: Evolution of phase diagrams and chaos including spin-glass order in d=2, Phys. Rev. E {\bf 89}, 042139 (2014).

\bibitem{Artun} E. C. Artun and A. N. Berker, Complete density calculations of q-state Potts and clock models: Reentrance of interface densities under symmetry breaking, arXiv:2005.00474 [cond-mat.stat-mech] (2020).

\bibitem{McCoyWu} B. M. McCoy and T. T. Wu, \textit{The Two-Dimensional Ising Model}, (Harvard University Press, Cambridge, 1973).

\bibitem{3dIsing} K. Binder and E. Luijten, Monte Carlo test of renormalization-group predictions for critical phenomena in Ising models, Phys. Rep. {\bf 344}, 179 (2001).

\bibitem{Pelissetto} A. Pelissetto abd E. Vicari, Critical phenomena and renormalization-group theory, Phys. Rep. {\bf 368}, 549 (2002).

\bibitem{Yesilleten} D. Ye\c{s}illeten and A. N. Berker, Renormalization-group calculation of local magnetizations and
correlations: Random-bond, random-field, and spin-glass systems,
Phys. Rev. Lett. {\bf 78}, 1564 (1997).

\bibitem{Fernandez2} A. Billoire, L. A. Fernandez, A. Maiorano, E. Marinari, V. Martin-Mayor, J. Moreno-Gordo, G. Parisi, F. Ricci-Tersenghi, J.J. Ruiz-Lorenzo, Dynamic variational study of chaos: Spin glasses in three dimensions, J. Stat. Mech. - Theory and Experiment, 033302 (2018).

\bibitem{Eldan} R. Eldan, The Sherrington-Kirkpatrick spin glass exhibits chaos, arXiv:2004.14885 (2020)

\end{references}
\end{document}